\documentstyle[12pt]{article}
\begin{document}
\title{The $s - d$ Transition in Heavy Mesons}
\author{Da-Xin Zhang\\
\it\small Institute of Theoretical Physics,\\[-3mm]
\it\small
School of Physics, Peking University,
Beijing 100871, China\footnote{Mailing address}
\\[-3mm]
\it\small and\\[-3mm]
\it\small
Institute of Theoretical Physics, Academia Sinica,\\ [-3mm]
\it\small P.O.Box 2735, Beijing 100080, China
}

\maketitle

\begin{abstract}
A  class of $|\Delta s|=1$ transition
is analyzed in the $D_s^+-D^+$ and $\bar B_s^0-\bar B^0$ systems.
Short distance Wilson coefficients are calculated within HQET.
Novel features of the transitions are discussed.
We find that these transitions are unobservable in the standard model.
\end{abstract}
\vspace{0.5cm}
{\it PACS}: 12.15.Mm, 13.90.+i, 14.40.Lb, 14.40.Nd

%\newpage
\section{Introduction}

The $\Delta I=1/2$ rule \cite{dI}  in the
$|\Delta s|=1$ decays of the kaons and the hyperons
is a great puzzle for the physicists to explain.
Theoretically,
using the techniques of
operator product expansion and renormalization group equations,
one gets the effective hamiltonian of
a set of local four-quark operators whose Wilson coefficients
contain the information at high energy.
There are big uncertainties in the
hadronic matrix elements of these four-quark operators
because the calculations are highly model dependent \cite{wu}.
Thus a comparison between the data and the prediction
 is not conclusive
in testing the standard model.
The $\Delta I=1/2$ rule is attributed
either to the uncertainties in
the hadronic matrix elements \cite{wu} or
to the possible existence of
new physics beyond the standard model \cite{kagan}.

We note that there
are more transitions induced by the $|\Delta s|=1$
effective hamiltonian other than the kaon and the hyperon decays.
In the standard model,
the effective hamiltonian for the $|\Delta s|=1$ transitions
including QCD-penguins is \cite{buras}
\begin{equation}
H_{eff}= \frac{G_F}{\sqrt{2}} V_{ud}^* V_{us}
\left[ \sum_{i=1}^{2} C_i(\mu) \left(Q_i^u-Q_i^c\right)
+\sum_{i=3}^{6}  C_i(\mu) \sum_{q=uds}^{cb} Q_i^q\right]
\label{heff0}
\end{equation}
between the $m_W$ and the $m_b$ scale,
where
\begin{eqnarray}
Q_{1}^u & = & \left( \bar d s  \right)_{\rm V-A}
            \left( \bar u u \right)_{\rm V-A}, \nonumber \\
Q_{2}^u & = & \left( \bar d_i s_j \right)_{\rm V-A}
            \left( \bar u_j u_i\right)_{\rm V-A}\nonumber \\
Q_{3}^q & = & \left( \bar d s\right)_{\rm V-A}
   \left( \bar q q \right)_{\rm V-A}, \nonumber \\
Q_{4}^q & = & \left( \bar d_{i} s_{j}  \right)_{\rm V-A}
   \left( \bar q_{j}  q_{i} \right)_{\rm V-A}\nonumber \\
Q_{5}^q & = & \left( \bar d s \right)_{\rm V-A}
   \left( \bar q q \right)_{\rm V+A}, \nonumber \\
Q_{6}^q & = & \left( \bar d_{i} s_{j}  \right)_{\rm V-A}
   \left( \bar q_{j}  q_{i} \right)_{\rm V+A}. \label{oper}
\end{eqnarray}
As the decay processes are concerned,
the operators
\begin{eqnarray}
Q_1^c &=& \left(\bar d s  \right)_{\rm V-A}
        \left(\bar c c \right)_{\rm V-A},\nonumber\\
Q_2^c &=& \left(\bar d_i s_j \right)_{\rm V-A}
        \left(\bar c_j c_i \right)_{\rm V-A} \label{operc}
\end{eqnarray}
and
\begin{eqnarray}
Q_{3\dots 6}^{c,b}
\label{opercb}
\end{eqnarray}
in (\ref{heff0}) are integrated out at the mass scale $m_c (m_b)$.

As the standard model is assumed,
the Wilson coefficients
for the operators with $q=c,b$ and $q=u,d,s$ in (\ref{heff0})
are related.
We will not consider the electroweak penguins whose inclusion is
straightforward and less important.
Although the operators in (\ref{operc})
and in (\ref{opercb})
do not contribute to the decay processes of
the kaon and the hyperons,
they could be relevant in other interesting processes in the
heavy hadron sectors.
$Q_{i=1\dots 6}^c$
could contribute to the $D_s^+-D^+$ mixing,
while $Q_{i=3\dots 6}^b$ could contribute
to the $\bar B_s^0-\bar B^0$ mixing,
if suitable modification by going into the
heavy quark effective theory (HQET) \cite{hqet}
where only the heavy degrees of
freedom are integrated out.

In Section 2 we present the novel features of the transition.
The Wilson coefficients and the hadronic matrix elements
of the relevant
operators in the standard model are calculated in Section 3
and Section 4, respectively.
We discuss the result in Section 5.

\section{Novel features of the transition}

To make a naive estimation for the amplitude of the
$\bar B_s^0-\bar B^0$ mixing induced by the penguin operators
$Q_{i=3\dots 6}^b$,
we compare these three groups of diagrams:
\begin{enumerate}
\item box diagrams $(b\bar d)-(d\bar b)$, $(b\bar s)-(d\bar b)$
and $(b\bar s)-(s\bar b)$;
\item box diagrams $(b\bar d)-(s\bar b)$ and $(b\bar d)-(b\bar s)$;
\item box diagram $(b\bar d)-(b\bar s)$ and penguin diagram
$(b\bar d)-(b\bar s)$.
\end{enumerate}
In the first group of diagrams,
the first and the third box diagrams are responsible for the
$B_d-\bar B_d$ and $B_s-\bar B_s$ mixings, respectively.
The second diagram in the first group is drawn by
replacing the $s$-quark by $d$-quark in the third diagram.
It induces a $|\Delta B|=2$ but $|\Delta s|=1$ mixing
$\bar B_s-B_d$,
which is smaller than the $B_s-\bar B_s$ mixing
in amplitude by a factor $\displaystyle\frac{V_{td}}{V_{ts}}$
but is bigger than the $B_d-\bar B_d$ mixing
by a factor $\displaystyle\frac{V_{ts}}{V_{td}}$.
In the second group,
the second diagram is drawn by exchanging the
external lines $s\to b$ and $\bar b\to \bar s$
in the first diagram.
Note that in the box diagrams the internal quark lines
run over $u,c,t$.
If the top quark gives the dominant
contribution to the second diagram,
then these two diagrams differ in amplitude by a factor
$\displaystyle\frac{V_{ts}^*V_{tb}}{V_{ts}V_{tb}^*}$
which is purely a phase factor.
We know that in the case of $K\bar K$ mixing,
the power suppressed contribution from the charm quark
is larger than highly CKM-suppressed
contribution from the top quark,
thus the second box diagram is analogously larger than the
first diagram in the second group.
In the third group,
the QCD-penguin diagram is approximately larger than
the box diagram by a
factor $\displaystyle\frac{\alpha_s}{\alpha_2}$ in amplitude.
We arrive at that the  penguin induced
$\bar B_s^0-\bar B^0$ mixing is comparable with
$B_s-\bar B_s$ mixing in amplitude.
The amplitude for $D_s^+-D^+$ mixing is even larger since the
tree-diagrams are not GIM suppressed.

Differing from the well known mechanism for
the neutral meson mixings of
$K^0-\bar K^0$, $D^0-\bar D^0$ and $B_{d,s}^0-\bar B_{d,s}^0$,
the cases of $D_s^+-D^+$ and $\bar B_s^0-\bar B^0$ mixings
have some novel features.
The system under consideration is non-degenerate states.
Both the mass difference
$\delta^D\equiv m_{D_s}-m_{D_d}$ and $\delta^B\equiv m_{B_s}-m_{B_d}$
are of the order of  100 MeV \cite{pdg},
comparing to the very small upper limit in the $K^0-\bar K^0$
due to the CPT invariance.
Defining
\begin{eqnarray}
\Delta^D &\equiv& \frac{1}{2 m_D}<D_s^+|H_{eff}|D^+>, \nonumber \\
\Delta^B &\equiv& \frac{1}{2 m_B}<\bar B_s^0|H_{eff}|\bar B^0>,
\label{DB}
\end{eqnarray}
the mass matrices of the $D_s-D_d$ and $B_s-B_d$ are
\begin{eqnarray}
\hat M_D=
\left(\begin{array}{cc}
m_{D_s} & \Delta^D\\
\Delta^{D\dagger} & m_D
\end{array}\right),
~~
\hat M_B=
\left(\begin{array}{cc}
m_{B_s} & \Delta^B\\
\Delta^{B\dagger} & m_B
\end{array}\right).
\label{mass}
\end{eqnarray}
The mass shifts are
$\pm \displaystyle\frac{\left|\Delta^{D,B}\right|^2}{\delta^{D,B}}$,
due to the see-saw mechanism.
The additional  factors
$\displaystyle\frac{\Delta^{D,B}}{\delta^{D,B}}$
suppress the mass shifts strongly to be unobservable,
although the transition amplitudes $\Delta^{D,B}$ can be much
larger than $\Delta m_K$ {\it etc}.

In the presence of $s-d$ transition in the heavy meson sector,
we need to view the observed states $D^+(1869)$ and
$\bar B^0(5279)$ as mass eigenstates which contain small amounts of
valence $\bar s$-quark.
Similar observation applies to
$D_s^+(1968)$ and $\bar B_s^0(5369)$.
Consequently,
at the B-factories a fraction
$\left|\displaystyle\frac{\Delta^{D,B}}{\delta^{D,B}}\right|^2$
of the neutral $\bar B^0(5279)$ mesons
decay as $\bar B_s$,
which have typical channel like
\begin{equation}
\bar B^0(5279)\rightarrow (b\bar s)\rightarrow
\left(D_s^+(1968),D_s^{+*}(2112)\right)+l\bar \nu.
\label{mode}
\end{equation}
Note that the semileptonic branching ratio of $\bar B_s$ into
$D_s^{+(*)}$ is about 10\% \cite{pdg},
these channels can be observed
if we have good mass-reconstruction of the final states.
Observation of the $D_s^+-D^+$ mixing effects
need to assume ideal mixing of $\rho-\omega-\phi$ so
that $\phi$ is a pure $s\bar s$ state,
then
\begin{equation}
D^+(1869)\rightarrow (c\bar s)\rightarrow
\phi \bar l \nu
\label{Dmode}
\end{equation}
is a characteristic channel.

\section{The short distance analysis}

In the HQET, the effective hamiltonian
below the $m_Q$ scale is
\begin{equation}
H_{eff} = \frac{G_F}{\sqrt{2}} V_{us}^* V_{ud}
\sum_{i=1}^{4} C_i^Q (\mu) O_i^Q,
\label{heffQ}
\end{equation}
where
\begin{eqnarray}
O_{1}^Q & = & \left( \bar d s \right)_{\rm V-A}
   \left( \bar h^{(Q)} h^{(Q)}\right)_{\rm V-A}, \nonumber \\
O_{2}^Q & = & \left( \bar d_{i} s_{j}  \right)_{\rm V-A}
   \left( \bar h^{(Q)}_{j}  h^{(Q)}_{i} \right)_{\rm V-A}\nonumber \\
O_{3}^Q & = & \left( \bar d s \right)_{\rm V-A}
   \left( \bar h^{(Q)} h^{(Q)} \right)_{\rm V+A},  \nonumber \\
O_{4}^Q & = & \left( \bar d_{i} s_{j}  \right)_{\rm V-A}
   \left( \bar h^{(Q)}_{j}  h^{(Q)}_{i} \right)_{\rm V+A}.
\label{operQ}
\end{eqnarray}
$h^{(Q)} (Q=c,b)$ is the heavy quark field in the HQET.

In the case of $D_s^+-D^+$ mixing all these $Q_1$ to $Q_6$
contribute to the matching at $\mu=m_c$
\begin{eqnarray}
C_1^c(m_c) &=& -C_1(m_c)+C_3(m_c),\nonumber\\
C_2^c(m_c) &=& -C_2(m_c)+C_4(m_c),\nonumber\\
C_3^c(m_c) &=& C_5(m_c),\nonumber\\
C_4^c(m_c) &=& C_6(m_c),\label{matchc}
\end{eqnarray}
and for $\bar B_s^0-\bar B^0$ mixing
only penguin operators contribute
\begin{eqnarray}
C_1^b(m_b) &=& C_3(m_b),\nonumber\\
C_2^b(m_b) &=& C_4(m_b),\nonumber\\
C_3^b(m_b) &=& C_5(m_b),\nonumber\\
C_4^b(m_b) &=& C_6(m_b).\label{matchb}
\end{eqnarray}

We now calculate the anomalous dimension matrix
in the HQET for the operators $O_i^Q$ and find
\begin{eqnarray}
\hat \gamma=
\left(\begin{array}{cc}
\hat\gamma_2 & 0\\ 0 &\hat\gamma_2
\end{array}\right),
~~\hat\gamma_2=\displaystyle\frac{\alpha_s}{4\pi}
\left(\begin{array}{cc}
0 & 0\\ 3 & -9
\end{array}\right).\label{anom}
\end{eqnarray}
Using the renormalization group equations
\begin{eqnarray}
\mu\frac{{\rm d}}{{\rm d}\mu}\hat C^Q=\hat \gamma^T C^Q
\label{rge}
\end{eqnarray}
the Wilson coefficients $C_i^Q(\mu)$ are then running
to the hadronic scale where the hadronic matrix
elements of $O_i^Q$ are to be calculated.
We have
\begin{eqnarray}
C_{1(3)}^c(\mu)&=&C_{1(3)}^c(m_c)-\frac{1}{3}
\left[\displaystyle
\left(\frac{\alpha_s(\mu)}{\alpha_s{(m_c)}}\right)^{27/58}
-1\right] C_{2(4)}^c(m_c), \nonumber\\
C_{2(4)}^c(\mu)&=&
\displaystyle
\left(\frac{\alpha_s(\mu)}{\alpha_s{(m_c)}}\right)^{27/58}
C_{2(4)}^c(m_c) \label{wilsonc}
\end{eqnarray}
for the $D_s^+-D^+$ mixing,
and
\begin{eqnarray}
C_{1(3)}^b(\mu)&=&C_{1(3)}^b(m_b)-\frac{1}{3}
\displaystyle
\left(\frac{\alpha_s(m_c)}{\alpha_s{(m_b)}}\right)^{27/58}
\left[\displaystyle
\left(\frac{\alpha_s(\mu)}{\alpha_s{(m_c)}}\right)^{9/20}-1
\right]
C_{2(4)}^b(m_b), \nonumber\\
C_{2(4)}^b(\mu)&=&
\displaystyle
\left(\frac{\alpha_s(m_c)}{\alpha_s{(m_b)}}\right)^{27/58}
\displaystyle
\left(\frac{\alpha_s(\mu)}{\alpha_s{(m_c)}}\right)^{9/20}
C_{2(4)}^b(m_b) \label{wilsonb}
\end{eqnarray}
for the $\bar B_s^0-\bar B^0$ mixing.
The scale independence of $3C_1+C_2$ and $3C_3+C_4$ follows simply
the form of the anomalous dimension (\ref{anom}).

\section{The hadronic matrix elements}

For the heavy meson mixings,
the matrix elements $<D_s^+|O_i^c|D^+>$
and $<\bar B_s^0|O_i^b|\bar B^0>$
are  much easier to calculate without
the notoriously difficulties in
the hadronic matrix elements of $Q_i$ for the $s$-quark
decay processes.
% Furthermore,
% in the decay processes there is a gap in scales between running the
% Wilson coefficients and calculating the hadronic matrix elements.
% This problem is absent in the case of heavy meson mixings.
Lacking lattice calculations of these matrix elements at present,
we use SU(3) symmetry to relate them to
those matrix elements of four-quark operators
$<\bar B_{(s)}^0|(\bar q q)(\bar h^Q h^Q)|\bar B_{(s)}^0>$
and
$<\bar B_{(s)}^0|(\bar q_i q_j)(\bar h^Q_j h^Q_i)|\bar B_{(s)}^0>$
and get
\begin{equation}
\begin{array}{rcccl}
<\bar B_s^0|O_1^b|\bar B^0>&=&<\bar B_s^0|O_3^b|\bar B^0>&\simeq&
\displaystyle\frac{1}{3}B_1f_B^2M_B^2,\nonumber\\
<\bar B_s^0|O_2^b|\bar B^0>&=&<\bar B_s^0|O_4^b|\bar B^0>&\simeq&
B_1f_B^2M_B^2
\label{4q}
\end{array}
\end{equation}
and similarly for $<D_s^+|O_i^c|D^+>$,
with $B_1\simeq 1$ calculated by QCD sum rules \cite{liu}.

\section{Results and discussions}

Within the HQET,
the Wilson coefficients are calculated at a scale
$\mu_0$ where $\alpha_s({\mu_0})\simeq 1$
to match the matrix elements\cite{wise}.
We also take $f_B=f_D=200$MeV in estimations.
We get
\begin{eqnarray}
\Delta^D &=& 1.4\times 10^{-7}, \nonumber \\
\Delta^B &=& 2.8\times 10^{-8}.
\label{numbers}
\end{eqnarray}
Correspondingly,
the mixing effects is  $2\times 10^{-12}$ for the $D_s^+-D^+$  system
and is $5\times 10^{-14}$ for the $\bar B_s^0-\bar B^0$ system
in probabilities.
We find that these small effects
are unobservable in the standard model.

Could any new physics exist beyond the standard model,
the Wilson coefficients for the heavy quark operators
in (\ref{operc}) and (\ref{opercb})
are not necessary to be related to those which are
relevant to the kaon and the hyperon decays.
Furthermore,
new operators with other Lorentz structures
are possibly relevant to the processes discussed here.
To extract a bound of order ${\cal O}(0.5)$
on these Wilson coefficients at the hadronic scale,
$\bar B^0$ decay events of a number of $10^{13}$ will be the minimal
if the new physics which induces the $\bar B_s^0-\bar B_d^0$
transition is at
the scale of 1TeV,
provided these events are good reconstructed
and the semileptonic decay fraction is 10\%.
It will be very interesting to construct
such kind of new physics models
which will be relevant to the processes discussed here.

\section*{Acknowledgments}
This work was supported in part by the National Natural Science
Foundation of China (NSFC) under the grant No. 90103014
and by the Ministry of Education of China.
I thank Chun Liu,
Chuan Liu, C.-D. Lu and M.-Z. Yang for helpful discussions,
and J.-P. Ma for hospitality when I was visiting ITP.

\end{document}